\documentclass[twocolumn]{article}
\usepackage[T1]{fontenc}
\usepackage{amsmath}

\usepackage{txfonts}

\newcommand{\textchange}[1]{{#1}}

\title{Strong Security and Separated Code Constructions
for the Broadcast Channels with Confidential Messages}
\date{May 2011}
\author{%
Ryutaroh Matsumoto\thanks{%
Dept.~of Communications and Integrated Systems,
  Tokyo Institute of Technology,
  Ookayama 2--12--1, Meguro-ku, Tokyo,
  152--8550 Japan.}
 \and
Masahito Hayashi\thanks{%
Graduate School of Information Sciences, Tohoku University, Aoba-ku, Sendai, 980-8579, Japan
and
Centre for Quantum Technologies, National University of Singapore, 3 Science Drive 2, Singapore 117542.}
}

\newtheorem{definition}{Definition}
\newtheorem{proposition}[definition]{Proposition}
\newtheorem{theorem}[definition]{Theorem}
\newtheorem{corollary}[definition]{Corollary}
\newtheorem{remark}[definition]{Remark}

\allowdisplaybreaks

\begin{document}
\maketitle
\begin{abstract}
We show that the capacity region of the broadcast channel
with confidential messages does not change when the strong
security criterion is adopted instead of the weak security criterion
traditionally used.
We also show a construction method of coding for
the broadcast channel with confidential messages
by using an arbitrary given coding for the broadcast channel
with degraded message sets.
\end{abstract}

\section{Introduction}\label{sec1}
The information theoretic security attracts much attention
recently \cite{liang09},
because it offers security that does not depend on
a conjectured difficulty of some computational problem.
A classical problem in the information theoretic security
is the broadcast channel with confidential messages
(hereafter abbreviated as BCC)
first considered by Csisz\'ar and K\"orner \cite{csiszar78},
in which there is a single sender called Alice and
two receivers called Bob and Eve.
The problem in \cite{csiszar78} is a generalization of
the wiretap channel considered by Wyner \cite{wyner75}.
In the formulation in \cite{csiszar78},
Alice has a common messages destined for both Bob and Eve and
a private message destined solely for Bob.
The word ``confidential'' means that Alice wants to
prevent Eve from knowing much about the private message.
The coding in this situation has two goals, namely
error correction and secrecy.

The traditional criterion of judging the secrecy
is the so-called weak security criterion,
which requires that the mutual information divided by
the code length between the secret message and the
adversary's information converges to zero as the
code length goes to the infinity.
Suppose that the secret message $S_n$ and the
adversary's message $Z^n$ of length $n$
have the identical part of length $n/\log n$ and
that the rests are statistically independent,
then the weak security criterion judges this situation
as secure, while the adversary knows infinitely much information
on the secret message.
This example suggests that the weak security criterion
is inappropriate for some applications.

To exclude such an insecure situation,
Maurer \cite{maurer94} introduced the
strong security criterion, which requires the
mutual information converges to zero without
division by the code length.
It is important to study the capacities
and the capacity regions of various information theoretical
problems under the strong security criterion.
The key agreement problem \cite{csiszar96,hayashi11,maurer00} and
the wiretap channels \cite{barros08,csiszar96,hayashi06b,hayashi11}
have been studied under the strong security criterion.
However, the capacity region for the BCC has not been
clarified as far as the authors' knowledge,
because the strong security results in \cite{barros08,csiszar96,hayashi06b,hayashi11}
for the wiretap channels
do not seem to easily generalize to the BCCs.
Note that \cite{maurer00} cannot be used to prove the strong
security of transmission of secret messages,
and that \cite{barros08} is an adaptation of \cite{maurer00}
to the wiretap channels.
We shall clarify that the capacity region under the
strong security criterion is the same as that under the
weak one.
Our proof argument just attaches inverses of hash functions
to an existing random coding argument for the
broadcast channel with degraded message sets (hereafter
abbreviated as BCD).
Thus the analysis of decoding error probability
in our proof automatically becomes as good as the
best analysis for BCD.
The idea of attaching inverses of hash functions seems
first appeared in Csisz\'ar \cite{csiszar96} in the
context of information theoretic security.

On the other hand,
in a communication system with single sender and single receiver,
the source coding and the channel coding are the most classical
and fundamental problems.
The famous source-channel separation theorem \cite[Section 7.13]{cover06}
states that we can get an optimal source-channel joint
coding by combining an optimal source coding and
an optimal channel coding, at least in the sense of
asymptotic information rate.
Therefore, it is natural to ask if there is
a similar separation theorem between secrecy coding and
error correction coding in the information theoretic security.
In this direction,
recently Csisz\'ar and Narayan \cite[Lemma B.2]{csiszar04}
and Renner \cite[Lemma 6.4.1]{rennerphd}
\emph{implicitly} proved the separation theorem between
secrecy coding and error correction coding in the
key agreement problem considered by Maurer \cite{maurer93} and
Ahlswede-Csisz\'ar \cite{ahlswede93}, which is also a classical
and fundamental problem in the information theoretic security.
Specifically, Csisz\'ar, Narayan, and Renner showed that
the optimal key rate can be attained for the model SW in \cite{ahlswede93}
with one-way public communication provided that we are given
optimal
probability distributions of the auxiliary random variables in
the key capacity formula,
by combining Slepian-Wolf encoder and decoder for
error correction and a family of two-universal hash functions
(fully random functions in \cite{csiszar04})
for secrecy.
However, the separation in other problems does not seem to be
explored, as far as the authors know.

Although our argument for the capacity region of
BCC in Section \ref{sec3}
separates the analysis of
the decoding error probability and the mutual information,
it does not separate the construction of a code for
error correction and provision of secrecy.
In Section \ref{sec4}
we introduce another form of the privacy amplification theorem
so that we can separate the code constructions for error
correction and secrecy,
then we clarify which rate pairs can be
achieved by our separated code construction.

This paper is organized as follows:
Section \ref{sec2} reviews relevant research results
used in this paper.
Section \ref{sec3} proves the capacity region under
the strong security criterion is the same as under the
weak one.
Section \ref{sec4} presents a computational procedure of
an upper bound on the mutual information when inverses
of hash functions are attached to an arbitrary given
code for BCD.
Section \ref{sec5} concludes the paper.

\section{Preliminaries}\label{sec2}
\subsection{Broadcast channels with confidential messages}
Let Alice, Bob, and Eve be as defined in Section \ref{sec1}.
$\mathcal{X}$ denotes the channel input alphabet and
$\mathcal{Y}$ (resp.\ $\mathcal{Z}$) denotes
the channel output alphabet to Bob (resp.\ Eve).
We assume that $\mathcal{X}$, $\mathcal{Y}$, and
$\mathcal{Z}$ are finite unless otherwise stated.
We shall discuss the continuous channel briefly in Remarks
\ref{rem:cont} and \ref{rem:cont2}.
We denote the conditional probability of the channel
to Bob (resp.\ Eve) by $P_{Y|X}$ (resp.\ $P_{Z|X}$).
The set $\mathcal{S}_n$ denotes that of the private message
and $\mathcal{E}_n$ does that of the common message
when the block coding of length $n$ is used.
We shall define the achievability of a rate triple
$(R_1$, $R_e$, $R_0)$,
where $R_1$ is the rate of the secret message,
$R_e$ is the so-called equivocation rate \cite{csiszar78},
and $R_0$ is the rate of the common message.
For the notational convenience, we fix the base of logarithm,
including one used in entropy and mutual information, to the
base of natural logarithm.
Privacy amplification theorems reviewed later are sensitive to
choice of the base of logarithm.

\begin{definition}
The rate triple $(R_1$, $R_e$, $R_0)$ is said
to be \emph{achievable} if there exists a sequence of
Alice's stochastic encoder $f_n$ from 
$\mathcal{S}_n \times \mathcal{E}_n$ to $\mathcal{X}^n$,
Bob's deterministic decoder $\varphi_n: \mathcal{Y}^n
\rightarrow \mathcal{S}_n \times \mathcal{E}_n$ and
Eve's deterministic decoder $\psi_n: \mathcal{Z}^n
\rightarrow \mathcal{E}_n$ such that
\begin{eqnarray*}
\lim_{n\rightarrow \infty} \mathrm{Pr}[
(S_n, E_n) \neq \varphi_n(Y^n) \textrm{ or } E_n \neq \psi_n(Z^n)]
&=& 0,\\
\liminf_{n\rightarrow \infty} \frac{H(S_n|Z^n)}{n} &\geq& R_e,\\
\liminf_{n\rightarrow\infty} \frac{\log|\mathcal{S}_n|}{n} &\geq& R_1,\\
\liminf_{n\rightarrow\infty} \frac{\log|\mathcal{E}_n|}{n} &\geq& R_0,
\end{eqnarray*}
where $S_n$ and $E_n$ represent the secret and the common message,
respectively, have the uniform distribution on $\mathcal{S}_n$
and $\mathcal{E}_n$, respectively,
and $Y^n$ and $Z^n$ are the received signal by Bob and Eve,
respectively, with the transmitted signal $f_n(S_n,E_n)$
and the channel transition probabilities $P_{Y|X}$, $P_{Z|X}$.
The capacity region of the BCC is the closure of 
the achievable rate triples.
\end{definition}

\begin{theorem}\label{th1}\cite{csiszar78}
The capacity region for the BCC is given by
the set of $R_0$, $R_1$ and $R_e$ such that
there exists a Markov chain $U\rightarrow V \rightarrow X \rightarrow 
YZ$ and
\begin{eqnarray*}
R_1 + R_0 &\leq& I(V;Y|U)+\min[I(U;Y),I(U;Z)],\\
R_0 &\leq& \min[I(U;Y),I(U;Z)],\\
R_e & \leq & I(V;Y|U)-I(V;Z|U),\\
R_e & \leq &R_1.
\end{eqnarray*}
\end{theorem}

As described in \cite{liang09},
$U$ can be regarded as the common message,
$V$ the combination of the common and the private messages,
and $X$ the transmitted signal.

If we set $R_e = R_1$ then we have
$\lim_{n\rightarrow\infty} I(S_n;Z^n)/n =0$,
which is traditionally called \emph{perfect security},
because Eve knows little about $S_n$.
However, Maurer \cite{maurer94} and Csisz\'ar \cite{csiszar96}
observed that 
$\lim_{n\rightarrow\infty} I(S_n;Z^n) =0$ is a better
criterion for the secrecy of $S_n$ from Eve,
and this stronger requirement is called the strong security criterion,
while the traditional one is called the weak security criterion
recently.

\begin{corollary}\label{cor:bcc}\cite{csiszar78}
The notation is same as Theorem \ref{th1}.
If we require $R_e = R_1$, the capacity region for
$(R_0$, $R_1)$ is given by
the set of $R_0$ and $R_1$ such that
there exists a Markov chain $U\rightarrow V \rightarrow X \rightarrow 
YZ$ and
\begin{eqnarray*}
R_0 &\leq& \min[I(U;Y),I(U;Z)],\\
R_1 & \leq & I(V;Y|U)-I(V;Z|U).
\end{eqnarray*}
\end{corollary}

\subsection{Broadcast channels with degraded message sets}\label{sec:bcd}
If we set $R_e= 0$ in the BCC,
the secrecy requirement is removed from BCC, and
the coding problem is equivalent to
the broadcast channel with degraded message sets (abbreviated as
BCD) considered by K\"orner and Marton \cite{korner77}.
\begin{corollary}
The capacity region of the BCD is given by
the set of $R_0$ and $R'_1$ such that
there exists a Markov chain $U\rightarrow V = X \rightarrow
YZ$ and
\begin{eqnarray*}
R_0 &\leq& \min[I(U;Y),I(U;Z)],\\
R_0+R'_1 & \leq & I(V;Y|U)+\min[I(U;Y),I(U;Z)].
\end{eqnarray*}
\end{corollary}
Throughout this paper, the information rate of
the private message to Bob without secrecy requirement
is denoted by $R'_1$ instead of $R_1$, to emphasize the difference.
One of several typical proofs for the direct part of
BCD is as follows \cite{bergmans73}:
Given $P_{UV}$, $R_0$, $R'_1$,
we randomly choose $\exp(nR_0)$ codewords of length $n$
according to $P^n_U$, and for each created codeword $u^n$,
randomly choose $\exp(nR'_1)$ codewords of length $n$
according to $P^n_{V|U}(\cdot|u^n)$.
Over the constructed ensemble of codebooks,
we calculate the average decoding probability by
the joint typical decoding, or the maximum likelihood decoding, etc.

\subsection{Privacy amplification theorem}
We shall use a family of two-universal hash functions \cite{carter79} and
a privacy amplification theorem obtained by Hayashi \cite{hayashi11} based on
the work by Bennett et al.~\cite{bennett95privacy}. So we shall review them.

\begin{definition}
Let $\mathcal{F}$ be a set of functions from $\mathcal{S}_1$ to $\mathcal{S}_2$,
and $F$ the not necessarily uniform
random variable on $\mathcal{F}$. If for any $x_1 \neq x_2
\in \mathcal{S}_1$ we have
\[
\mathrm{Pr}[F(x_1)=F(x_2)] \leq \frac{1}{|\mathcal{S}_2|},
\]
then $\mathcal{F}$ is said to be a \emph{family of two-universal hash functions}.
\end{definition}

\begin{proposition}\label{thm:hayashi09b}
Let $L$ be a random variable with a finite alphabet
$\mathcal{L}$ and $Z$ any random variable.
Let $\mathcal{F}$ be a family of two-universal
hash functions from $\mathcal{L}$ to $\mathcal{M}$,
and $F$ be a random variable on $\mathcal{F}$
statistically independent of $L$.
Then
\begin{equation}
I(F(L);Z|F)  \leq   
\frac{1}{\rho}
|\mathcal{M}|^\rho\mathbf{E}[P_{L|Z}(L|Z)^\rho]\label{hpa1}
\end{equation}
for $0<\rho\leq 1$.
If $Z$ is not discrete RV,
$I(F(L);Z|F)$ is defined to be
$H(F(L)|F) - \mathbf{E}_z H(F(L)|F,Z=z)$.

In addition to the above assumptions,
when $L$ is uniformly distributed, we have
\begin{equation}
\frac{1}{\rho}
|\mathcal{M}|^\rho\mathbf{E}[P_{L|Z}(L|Z)^\rho]=
\frac{|\mathcal{M}|^\rho\mathbf{E}[P_{L|Z}(L|Z)^\rho P_L(L)^{-\rho}]}{|\mathcal{L}|^\rho\rho}.\label{hpa1uni}
\end{equation}
In addition to all of the above assumptions,
when $Z$ is a discrete random variable, 
we have
\begin{eqnarray}
&&\frac{|\mathcal{M}|^\rho\mathbf{E}[P_{L|Z}(L|Z)^\rho P_L(L)^{-\rho}]}{|\mathcal{L}|^\rho\rho}\nonumber\\
&=&
\frac{|\mathcal{M}|^\rho}{|\mathcal{L}|^\rho\rho}
\sum_z \sum_\ell P_L(\ell) P_{Z|L}(z|\ell)^{1+\rho} P_Z(z)^{-\rho}.
\label{hpa1discrete}
\end{eqnarray}
\end{proposition}

\begin{remark}
It was assumed that $Z$ was discrete in \cite{hayashi11}.
However, when the alphabet of $L$ is finite,
there is no difficulty to extend the original result.
\end{remark}


As in \cite{hayashi11} we introduce the following two functions.
\begin{definition}
\begin{eqnarray}
\psi(\rho, P_{Z|L}, P_L) &=& 
\log \sum_z \sum_\ell P_L(\ell) P_{Z|L}(z|\ell)^{1+\rho} P_Z(z)^{-\rho},
\label{eq:psid}\\
\phi(\rho,P_{Z|L},P_L) 
&=& \log \sum_z\left(
\sum_{\ell} P_{L}(\ell) (P_{Z|L}(z|\ell)^{1/(1-\rho)})\right)^{1-\rho}.
\label{phid}
\end{eqnarray}
\end{definition}
Observe that $\phi$ is essentially Gallager's function $E_0$
\cite{gallager68}.

\begin{proposition}\cite{gallager68,hayashi11}
\textchange{$\exp(\phi(\rho, P_{Z|L}, P_L))$ is} concave with respect to $P_L$
with fixed $0<\rho< 1$ and $P_{Z|L}$.
For fixed $0<\rho< 1$, $P_L$ and $P_{Z|L}$ we have
\begin{equation}
\exp(\psi(\rho, P_{Z|L}, P_L))\leq \exp(\phi(\rho, P_{Z|L}, P_L)).
\label{psileqphi}
\end{equation}
\end{proposition}

\section{Calculation of the average mutual information with random coding}\label{sec3}
In this section we shall prove that the capacity region
given in Corollary \ref{cor:bcc} is also the capacity region
of the BCC under the strong security criterion.
We do not need the proof for the converse part.
We shall prove the direct part.
Let the RV $B_n$ on $\mathcal{B}_n$ denote the private
message to Bob \emph{without secrecy requirement},
$E_n$ on $\mathcal{E}_n$ the common message to
both Bob and Eve,
$F_n$ on $\mathcal{F}_n$ a function in a family $\mathcal{F}_n$
of two-universal hash functions from $\mathcal{B}_n$ to
$\mathcal{S}_n$, $\Lambda$ an RV indicating selection of
codebook in the random ensemble constructed in the way
reviewed in Section \ref{sec:bcd},
$U^n = \Lambda(E_n)$ on $\mathcal{U}^n$
and $V^n=\Lambda(B_n,E_n)$ on $\mathcal{V}^n$ codewords
for the BCD taking the random selection $\Lambda$ taking
into account, and $Z^n$ Eve's received signal,
where $n$ denotes the code length.
We assume that for every $f_n \in \mathcal{F}_n$ is surjective
and for all $s \in \mathcal{S}_n$ the set $\{
b \in \mathcal{B}_n \mid f_n(b) = s\}$ has the constant
number of elements.
Such requirement on $f_n$ is satisfied, for example, when
$\mathcal{F}_n$ is the set of all surjective linear maps
from $\mathcal{B}_n$ to $\mathcal{S}_n$.

The structure of the transmitter and the receiver is as follows:
Fix a hash function $f_n \in \mathcal{F}_n$ and
Alice and Bob agree on the choice of $f_n$.
Given a secret message $s_n$,
choose $b_n$ uniformly randomly from $\{ b \in \mathcal{B}_n \mid
f_n(b) = s_n \}$, treat $b_n$ as the private message to Bob,
encode $b_n$ along with the common message $e_n$ by
an encoder for the BCD, and get a codeword $v^n$.
Apply the artificial noise to $v^n$ according to the
conditional probability distribution $P^n_{X|V}$ and
get the transmitted signal $x^n$.
Bob decodes the received signal and get $b_n$, then
apply $f_n$ to $b_n$ to get $s_n$.
This construction requires Alice and Bob to agree on the
choice of $f_n$.
We shall show that $I(S_n; Z^n|F_n) = \mathbf{E}_{f_n}
I(S_n; Z^n|F_n=f_n)$ to be arbitrary small.
This ensures that most choice of $f_n$ makes
$I(S_n; Z^n|F_n=f_n)$ small.
The same argument was also used in \cite{csiszar96}.

Let $S_n$ denote the RV of the secret message.
Define $B'_n$ to be the RV uniformly chosen from the
random set $\{ b \in \mathcal{B}_n \mid
F_n(b) = S_n \}$.
We want to apply the privacy amplification theorem
to $I(F_n(B'_n);Z^n|F_n)$.
To use the theorem (Proposition \ref{thm:hayashi09b}) we must ensure
independence\footnote{The statistical independence of the
corresponding random variables in \cite{barros08,csiszar96}
was not discussed in detail.} of $F_n$ and $B'_n$.
The independence is satisfied by the assumptions on
$\mathcal{F}_n$ if $S_n$ is uniformly distributed.
In that case $B'_n$ is uniformly distributed over
$\mathcal{B}_n$. Denote $B'_n$ by $B_n$.
The remaining task is to find an upper bound
on $I(F_n(B_n);Z^n|F_n,\Lambda)$.
Since the decoding error probability of the above scheme
is not greater than that of the code for BCD,
we do not have to analyze the decoding error probability.

Firstly, we consider $I(F_n(B_n);Z^n|F_n,\Lambda)$ with
fixed selection $\lambda$ of $\Lambda$.
In the following analysis,
we do not make any assumption on the probability
distribution of $E_n$ except that $S_n$, $E_n$, $F_n$ and $\Lambda$
are statistically independent.

\begin{eqnarray}
&&I(F_n(B_n);Z^n|F_n,\Lambda=\lambda)\nonumber\\
&\leq &I(F_n(B_n);Z^n,E_n|F_n,\Lambda=\lambda)\nonumber\\
&=& \underbrace{I(F_n(B_n);E_n|F_n,\Lambda=\lambda)}_{=0}+
I(F_n(B_n);Z^n|F_n,E_n,\Lambda=\lambda)\nonumber\\
&=& \sum_{e}P_{E_n}(e)I(F_n(B_n);Z^n|F_n,E_n=e,\Lambda=\lambda)\label{hbefore}\\
&\leq&\sum_{e}P_{E_n}(e)\frac{\exp(n\rho R_1)}{\rho\exp(n\rho R'_1)} \sum_{b,z}P_{B_n}(b)\nonumber\\
&&P_{Z^n|B_n,E_n,\Lambda=\lambda}(z|b,e)^{1+\rho}
P_{Z^n|E_n=e,\Lambda=\lambda}(z)^{-\rho}\nonumber
\textrm{ (by Eqs.\ (\ref{hpa1}--\ref{hpa1discrete}))}\nonumber\\
&=&\sum_{e}P_{E_n}(e)\frac{\exp(n\rho R_1)}{\rho\exp(n\rho(R'_1))} \sum_{v,z}\underbrace{\sum_{b:\lambda(b,e)=v}P_{B_n}(b)}_{=P_{V^n|E_n=e,\Lambda=\lambda}(v)}\nonumber\\
&&\underbrace{P_{Z^n|B_n,E_n,\Lambda=\lambda}(z|b,e)^{1+\rho}}_{=P_{Z^n|V^n,\Lambda=\lambda}(z|v)^{1+\rho}}P_{Z^n|E_n=e,\Lambda=\lambda}(z)^{-\rho}\nonumber\\
&=&\sum_{e}P_{E_n}(e)\frac{\exp(n\rho R_1)}{\rho\exp(n\rho(R'_1))} \sum_{v,z}P_{V^n|E_n=e,\Lambda=\lambda}(v)\nonumber\\
&&P_{Z^n|V^n,\Lambda=\lambda}(z|v)^{1+\rho}
P_{Z^n|E_n=e,\Lambda=\lambda}(z)^{-\rho}\nonumber\\
&=&\sum_{e}P_{E_n}(e)\frac{\exp(n\rho R_1 + \psi(\rho,P_{Z^n|V^n,\Lambda=\lambda},P_{V^n|E_n=e,\Lambda=\lambda} ))}{\rho\exp(n\rho(R'_1))}\nonumber\\
&=&\sum_{e}P_{E_n}(e)\frac{\exp(n\rho R_1 + \psi(\rho,P_{Z^n|V^n},P_{V^n|E_n=e,\Lambda=\lambda} ))}{\rho\exp(n\rho(R'_1))}\nonumber\\
&=&\sum_{e}P_{E_n}(e)\frac{\exp(n\rho(R_1-R'_1) + \psi(\rho,P_{Z^n|V^n},P_{V^n|E_n=e,\Lambda=\lambda}))}{\rho}.\nonumber\\
&&\label{hafter}
\end{eqnarray}

We shall average the above upper bound over $\Lambda$.
\begin{eqnarray}
&&\sum_{e}P_{E_n}(e)I(F_n(B_n);Z^n|F_n,{\Lambda},E_n=e)\nonumber\\
&\leq&\sum_{\lambda}P_\Lambda(\lambda)\sum_{e}P_{E_n}(e)\nonumber\\*
&&\frac{\exp(n\rho(R_1-R'_1) + \psi(\rho,P_{Z^n|V^n},P_{V^n|{E_n=e},\Lambda=\lambda} ))}{\rho}\nonumber\\
&=&\sum_{\lambda}P_\Lambda(\lambda)\sum_{e}P_{E_n}(e)\nonumber\\*
&&\frac{\exp(n\rho(R_1-R'_1) + \psi(\rho,P_{Z^n|V^n},P_{V^n|{U^n=\lambda(e)},\Lambda=\lambda} ))}{\rho}\nonumber\\
&=&\sum_{\lambda}P_\Lambda(\lambda)\sum_{u}P_{U^n|\Lambda=\lambda}(u)\nonumber\\*
&&\frac{\exp(n\rho(R_1-R'_1) + \psi(\rho,P_{Z^n|V^n},P_{V^n|U^n=u,\Lambda=\lambda} ))}{\rho}\nonumber\\
&=&\sum_{u}P_{U^n}(u)\sum_{\lambda}P_{\Lambda|U^n=u}(\lambda)\nonumber\\*
&&\frac{\exp(n\rho(R_1-R'_1) + \psi(\rho,P_{Z^n|V^n},P_{V^n|U^n=u,\Lambda=\lambda} ))}{\rho}\nonumber\\
&\leq&\sum_{u}P_{U^n}(u)\sum_{\lambda}P_{\Lambda|U^n=u}(\lambda)\nonumber\\*
&&\frac{\exp(n\rho(R_1-R'_1) + \textchange{\phi}(\rho,P_{Z^n|V^n},P_{V^n|U^n=u,\Lambda=\lambda} ))}{\rho}\nonumber\\
&&\textrm{(by Eq.\ (\ref{psileqphi}))}\nonumber\\
&\leq&\frac{1}{\rho}\sum_{u}P_{U^n}(u)
\exp\Bigl[n\rho(R_1-R'_1) \nonumber\\*
&&+ \textchange{\phi}\Bigl(\rho,P_{Z^n|V^n},\sum_{\lambda}P_{\Lambda|U^n=u}(\lambda)P_{V^n|U^n=u,\Lambda=\lambda} \Bigr)\Bigr]\nonumber\\
&&\textrm{(concavity of $\exp(\textchange{\phi})$ is used)}\label{eq:useconcavity}\\
&=&\frac{1}{\rho}\sum_{u^n\in\mathcal{U}^n}P_{U^n}(u^n)\exp[n\rho(R_1-R'_1) \nonumber\\*
&&+ \textchange{\phi}(\rho,P_{Z^n|V^n},P_{V^n|U^n=u^n} )]\nonumber\\
&=&\frac{1}{\rho}\sum_{u^n\in\mathcal{U}^n}\prod_{i=1}^n P_{U}(u_i)\exp[\rho(R_1-R'_1) \nonumber\\*
&&+ \textchange{\phi}(\rho,P_{Z|V},P_{V|U=u_i} )]\nonumber\\
&=&\frac{1}{\rho}\prod_{i=1}^n \sum_{u_i\in\mathcal{U}}P_{U}(u_i)\exp[\rho(R_1-R'_1) \nonumber\\*
&&+ \textchange{\phi}(\rho,P_{Z|V},P_{V|U=u_i} )]\nonumber\\
&=&\frac{1}{\rho}\Bigl[
\exp(\rho(R_1-R'_1))\nonumber\\*
&&\sum_{u\in\mathcal{U}}P_{U}(u)\exp(\textchange{\phi}(\rho,P_{Z|V},P_{V|U=u} ))\Bigr]^n.\label{eq:final}
\end{eqnarray}

We shall consider
under what condition the upper bound goes to zero.
Taking the logarithm of the upper bound (\ref{eq:final}) we have
\begin{eqnarray*}
&&-\log \rho +n\rho \Biggl[R_1-R'_1 + \\
&&\underbrace{\frac{1}{\rho}\log
\Bigl(\sum_{u\in\mathcal{U}}P_{U}(u)\exp(\textchange{\phi}(\rho,P_{Z|V},P_{V|U=u} ))\Bigr)}_{(*)}\Biggr].
\end{eqnarray*}
\textchange{We can see that (*) $\rightarrow I(V;Z|U)$ as $\rho \rightarrow 0$.
by applying the l'H\^opital's rule to (*).}

This shows that the amount $R'_1-R_1$ of random garbage required to make
$S_n = F_n(B_n)$ secret from Eve is $I(V;Z|U)$ per channel use.
By choosing $R_0 = \min\{I(U;Y),I(U;Z)\}-\delta$ and $R'_1 =
I(V;Y|U)-\delta$, we have completed the direct part proof.
\hfill\rule{1ex}{1ex}

\begin{remark}
Our proof does not require the common message $E_n$ to be decoded
by Bob. Our technique can provide an upper bound on the mutual
information of $S_n = F_n(B_n)$
even when $E_n$ is a private message to Eve.
\end{remark}

\begin{remark}
The (negative) exponential decreasing rate of the mutual
information in our argument is
\[
\textchange{\rho \Biggl[R_1-R'_1 + 
\frac{1}{\rho}\log
\Bigl(\sum_{u\in\mathcal{U}}P_{U}(u)\exp(\textchange{\phi}(\rho,P_{Z|V},P_{V|U=u} ))\Bigr)\Biggr]}
\]
Minimizing the above expression over $0 < \rho \leq 1$,
$R'_1$ and
$U\rightarrow V\rightarrow X \rightarrow YZ$ such that
$R_0 \leq \min\{ I(U;Y)$, $I(U;Z)\}$,
and $R'_1 \leq I(V;Y|U)$
gives the smallest negative exponent. From
the form of the mathematical expression,
increase in $R'_1$ decreases the mutual information
and increases the decoding error probability of the secret
message to Bob. This suggests that the optimal mutual
information and the optimal decoding error probability
cannot be realized simultaneously.
\end{remark}

\begin{remark}\label{rem:cont}
We can easily carry over our proof to the case of
 the channel being Gaussian,
because
\begin{itemize}
\item we can extend Eq.\ (\ref{hpa1discrete}) to the Gaussian case
just by replacing the probability mass functions $P_{Z|L}$ and $P_Z$ by
their probability density functions.
\item the random codebook $\Lambda$ obeys the
multidimensional Gaussian distribution,
\item the concavity of $\textchange{\phi}$ is retained when
its second argument is conditional probability density,
\item and the all mathematical manipulations in this
section remains valid when
$U$, $V$, $Z$, $\Lambda$ are continuous and
their probability mass functions are replaced with
probability density functions,
while $B_n$, $E_n$, $F_n$ remain to be discrete RVs on
finite alphabets.
\end{itemize}
\end{remark}

\section{Separated code construction 
for the broadcast channel with confidential messages}\label{sec4}
Suppose that we are given single triple of an encoder and
decoders for BCD.
We want to construct a code for BCC based on the code for BCD,
by attaching the inverse of a randomly chosen two-universal hash function to
the given BCD code.
If we could do this without loss of any optimality,
then the practical study of codes for
BCC would become unnecessary,
because the study of practical BCC codes can be reduced to that
of practical BCD codes.
We stress that the random choice of encoder and decoder
is widely accepted as a practical method, see e.g., \cite{cai09,hayashi06,rennerphd}.

Let $X$ be the uniform distribution on the given codebook,
and $Z$ Eve's received signal given $X$ as channel input.
By simply applying Proposition \ref{thm:hayashi09b}
to $X$ and $Z$,
the size of secret message set $\mathcal{S}$ has to satisfy
\begin{equation}
\min_{0<\rho\leq 1}
\frac{\displaystyle |\mathcal{S}|^\rho \mathbf{E}[ P_{X|Z}(X|Z)^{\rho}]}{\rho}
\leq \textrm{acceptable value}. \label{eq:imp}
\end{equation}
When the number of codewords is, say $2^{1000}$, evaluation of
the left hand side is practically impossible.

We shall introduce another form of the privacy amplification theorem
alternative to Proposition \ref{thm:hayashi09b}
whose computation as Eq.\ (\ref{eq:imp})
is intractable with arbitrary given single BCD
encoder,
so that we can compute a suitable size of $\mathcal{S}$.
What follows is an extension of a  result on the
wiretap channel \cite{hayashimatsumoto10}.
The following theorem is an adaptation of
the channel resolvability lemma \cite[Lemma 2]{hayashi06b}.

\begin{theorem}\label{newresolvability2}
Assume that the given family of two-universal hash function $F$ from $\mathcal{L}$
to $\mathcal{M}$ satisfies that
\begin{eqnarray*}
|F^{-1}(m)|=\frac{|\mathcal{L}|}{|\mathcal{M}|}, \quad
\forall m,
\end{eqnarray*}
the statistically independent random variable $K$ and $L$ obey the uniform distributions on 
$\mathcal{K}$ and
$\mathcal{L}$, respectively, and
a fixed conditional probability $Q_{Z|K,L}$ is given.
We also assume that $F$ is statistically independent of $K$ and $L$.
Then,
\[
I(F(L); Z|F) \leq 
\frac{|\mathcal{M}|^\rho \exp(\phi(\rho,Q_{Z|K,L},P_{K,L}))}{(|\mathcal{K}|\times|\mathcal{L}|)^\rho \rho},
\]
for $0 < \rho < 1$.
\end{theorem}
\noindent\emph{Proof.}
\begin{eqnarray*}
&&
I(F(L); Z|F)\\
&\leq& I(F(L); K,Z|F)\\
&=& I(F(L); Z|K,F)\\
&\leq&
\sum_k P_K(k) \frac{|\mathcal{M}|^\rho}{|\mathcal{L}|^\rho\rho}
\exp(\psi(\rho,P_{Z|K=k,L},P_L)) \\
&&\textrm{ (by Eqs.\ (\ref{hpa1}--\ref{hpa1discrete}))}\\
&\leq&
\sum_k P_K(k) \frac{|\mathcal{M}|^\rho}{|\mathcal{L}|^\rho\rho}
\exp(\phi(\rho,P_{Z|K=k,L},P_L)) \textrm{ (by Eq.\ (\ref{psileqphi}))}\\
&=&
\sum_k \underbrace{P_K(k)P_K(k)^{\rho-1}}_{=|\mathcal{K}|^{-\rho}} \frac{|\mathcal{M}|^\rho}{|\mathcal{L}|^\rho\rho}
\sum_z\\
&&\left(
\sum_{\ell} P_K(k)P_{L}(\ell) (P_{Z|K,L}(z|k,\ell)^{1/(1-\rho)})\right)^{1-\rho}\\
&=&
\frac{|\mathcal{M}|^\rho}{|\mathcal{K}\times\mathcal{L}|^\rho\rho}
\sum_z\left(
\sum_{k,\ell} P_{K,L}(k,\ell) (P_{Z|K,L}(z|k,\ell)^{1/(1-\rho)})\right)^{1-\rho}\\
&=&
\frac{|\mathcal{M}|^\rho}{|\mathcal{K}\times\mathcal{L}|^\rho\rho}
\exp(\phi(\rho,P_{Z|K,L},P_{K,L})).
\end{eqnarray*}
\hfill\rule{1ex}{1ex}

Suppose that we are given a triple of an encoder and
decoders for the BCD. We shall derive a computable upper bound
on the mutual information between the secret message and
Eve's received signal.
Let $S_n$ be the secret message to Bob,
$E_n$ be the common message to both Bob and Eve.
We assume that $S_n$ and $E_n$ are uniformly distributed and that
they are statistically independent to each other.
Let $F_n$ be an RV on a family $\mathcal{F}_n$
of two-universal hash functions.
$F_n$ is statistically independent of $(S_n$, $E_n)$,
and we use the same assumptions on the hash functions
as Section \ref{sec3}.
Let $B_n$ be the uniform random on the set
$\{ b \in \mathcal{B}_n \mid F_n(b) = S_n \}$.
As in Section \ref{sec3}, $B_n$ and $F_n$ are statistically
independent and we can apply the privacy amplification theorem.
Let $Z^n$ be Eve's received signal after encoding $B_n$ by
the given encoder $e_n$
for BCD, applying the artificial noise
$P^n_{X|V}$, and transmitting the resulted signal over
the given channel.
By using Theorem \ref{newresolvability2}, we have
\begin{equation}
I(F_n(B_n);Z^n|F_n)
\leq 
\frac{|\mathcal{S}_n|^\rho \exp(\phi(\rho,P^n_{Z|V},P_{e_n(B_n,E_n)}))}{|\mathcal{B}_n\times\mathcal{E}_n|^\rho\rho}.\label{eq111}
\end{equation}
Recall that the function $\phi$ is essentially Gallager's function $E_0$ \cite{gallager68},
and we have
\begin{eqnarray}
&& \exp(\phi(\rho,P^n_{Z|V},P_{e_n(B_n,E_n)})) \nonumber\\
&\leq &
\max_{P_n\textrm{ on }\mathcal{V}^n} \exp(\phi(\rho,P^n_{Z|V},P_n))\nonumber\\
&=& \max_{P_1\textrm{ on }\mathcal{V}^1} \underbrace{\exp(n\phi(\rho,P^n_{Z|V},P_1))}_{(**)}
\textrm{ (by \cite{arimoto73})}. \label{singleletter}
\end{eqnarray}
Because (**) is concave with respect to $P_1$ \cite{gallager68},
its maximization can be computed
in practice, for example by \cite{boyd04}.
On the other hand,
$\min_{P_1}$(**) is convex with respect to $\rho$, so its minimization
with respect to  $\rho$ can also be computed
by the bisection method \cite{boyd04}.

We have to investigate under which conditions
the above computational procedure for the secret message
size can achieve a rate pair $(R_0$, $R_1)$.
The logarithm of the right hand side  of Eq.\ (\ref{eq111})
is
\[
\rho \left(\log |\mathcal{S}_n|-\log |\mathcal{B}_n|
- \log |\mathcal{E}_n| + \frac{n\phi(\rho,P_{Z|V},P_1)}{\rho}\right) - \log \rho. 
\]
Since $\phi$ is essentially $E_0$ in \cite{gallager68},
$\lim_{\rho\rightarrow 0} \phi(s,P_{Z|V},P_V)/\rho = I(V;Z)$.
Therefore,
if $\log|\mathcal{S}_n| < \log |\mathcal{B}_n| + \log |\mathcal{E}_n| - n (\max_{P_V}I(V;Z)+\delta)$
for all $n$, then $I(S_n;Z^n|F_n)$ goes to zero as $n\rightarrow \infty$.
This means that if $(R_0$, $R'_1)$ is achievable as a rate pair
in the BCD defined by $P_{YZ|V}$ by adjusting
the artificial noise $P_{X|V}$ and $R_1 \leq R_0 + R'_1 - \max_{P_V} I(V;Z)$
then the rate pair $(R_0$, $R_1)$ is achievable by our
computational procedure for an upper bound on the mutual
information.

When we use Theorem \ref{newresolvability2} in place of
Proposition \ref{thm:hayashi09b} in Section \ref{sec3},
we can prove an achievable rate pair $(R_0$, $R_1)$ to be
achievable by Theorem \ref{newresolvability2} only
if $R_0$ is close to $I(U;Z)$.
We cannot prove the achievability of a rate pair $(R_0$,
$R_1)$ by Theorem \ref{newresolvability2}
if $R_0 \simeq I(U;Y) < I(U;Z)$.

\begin{remark}\label{rem:cont2}
As Remark \ref{rem:cont},
the generalization of results in Section \ref{sec4}
to the Gaussian channels
is easy provided that
the transmitted signal is chosen from a fixed finite
subset of $\mathbf{R}$ for every channel use.
When the transmitted signal is chosen from varying
finite sets for each channel use,
we have difficulty in Eq.\ (\ref{singleletter}).
\end{remark}

\section{Conclusion}\label{sec5}
We argued that the weak security criterion,
which requires only the mutual information divided by the
code length converges to zero, may be inappropriate in
some applications, by explicitly providing an insecure example,
and made a case for the strong security criterion
introduced by Maurer \cite{maurer94},
which requires the mutual information converges to
zero without division by the code length.
The broadcast channel with confidential messages \cite{csiszar78}
is one
of fundamental problems in the information theoretical
security \cite{liang09},
but its capacity region remained unknown under the
strong security criterion before this paper.
We have shown that the capacity region
under the strong security is the same as that under the weak one.

On the other hand, the separation between secrecy coding and
error correction coding is important from both theoretical
and practical viewpoints. We presented a random coding argument
and a code construction that separate error correction and
secrecy. However, our separations for the broadcast channel with
confidential messages are still incomplete compared to the
source channel separation \cite[Section 7.13]{cover06} or
the separation in the classical \cite{csiszar04,rennerphd}
and quantum \cite{rennerphd} key agreement
problem,
because we cannot separately and independently
construct codes for secrecy and error correction and
combine them without losing the optimality in the sense
of asymptotic information rate, as done in
\cite[Section 7.13]{cover06} and \cite{csiszar04,rennerphd}.

\section*{Acknowledgment}
This research was in part conducted during the first author's stay
at the Institute of Network Coding, Chinese University of Hong
Kong. We appreciate the hospitality of Prof. Raymond W. Yeung and
members at INC.
We also thank Dr.\ Chung Chan for informing us about the separation of
coding in \cite{csiszar04} and the paper by Maurer and Wolf \cite{maurer00},
and Dr.\ Jun Muramatsu for helpful discussion on the universal coding.
This research was partially supported by 
the MEXT Grant-in-Aid for Young Scientists (A) No.\ 20686026
and No.\ 22760267.
The Center for Quantum Technologies is funded
by the Singapore Ministry of Education and the National Research
Foundation as part of the Research Centres of Excellence programme.


\end{document}